\documentclass[twocolumn,twocolappendix]{aastex631}
\usepackage{amsmath}
\usepackage{appendix}

\begin{document}

\title{Exomoons of Circumbinary Planets}

\author[0009-0000-3336-4335]{Ben R. Gordon}
\correspondingauthor{Benjamin.Gordon1@lsu.edu}
\affiliation{Department of Physics \& Astronomy, Tufts University, 574 Boston Avenue, Medford, MA 02155, USA}
\affiliation{Department of Physics \& Astronomy, Louisiana State University, Baton Rouge, LA 70803, USA}

\author[0009-0009-9909-8063]{Helena Buschermöhle}
\affiliation{Department of Physics \& Astronomy, Tufts University, 574 Boston Avenue, Medford, MA 02155, USA}

\author[0000-0002-7907-2634]{Wata Tubthong}
\affiliation{Department of Physics \& Astronomy, Tufts University, 574 Boston Avenue, Medford, MA 02155, USA}

\author[0000-0002-7595-6360]{David V. Martin}
\affiliation{Department of Physics \& Astronomy, Tufts University, 574 Boston Avenue, Medford, MA 02155, USA}

\author{Sean Smallets}
\affiliation{Department of Physics \& Astronomy, Tufts University, 574 Boston Avenue, Medford, MA 02155, USA}

\author{Grace Masiello}
\affiliation{Department of Physics \& Astronomy, Tufts University, 574 Boston Avenue, Medford, MA 02155, USA}

\author{Liz Bergeron}
\affiliation{Department of Physics \& Astronomy, Tufts University, 574 Boston Avenue, Medford, MA 02155, USA}



\begin{abstract}

Confirmation of the first exomoon remains elusive. Although several exomoon candidates exist around single stars, there are currently no candidates around circumbinary planets (CBPs). Most circumbinary planets are thought to form far from the host binary and migrate through the protoplanetary disc. Therefore, an exomoon of a CBP represents a fascinating yet complex and evolving four-body system. Their existence (or absence) would shed light on the robustness of moon formation and evolution in dynamically active planetary systems. In this work, we simulate the orbital evolutions of exomoons around migrating CBPs. We show that for fully migrated CBPs, a moon is capable of surviving the migration if it is formed within $\sim5-10\%$ of the planet's Hill Radius, well within the currently proposed range at which moons are thought to settle in the planetary disc for giant planets. Even though all known CBPs are gas giants, 18\% of the surviving moons in our sample are within the habitable zone, giving credence to circumbinary habitability, albeit hosted by moons rather than planets. $38\%$ of moons escape their host planet early in the migration and become long-period CBPs (i.e a multi-planet circumbinary system). Nearly one-third of exomoons collide with their host planet, and $1\%$ are ejected from the system entirely. This last class presents another pathway for producing free-floating planetary mass objects, like those discovered recently and expected from the Roman microlensing survey. 

\end{abstract}



\section{INTRODUCTION}\label{sec:introduction}

There are $>6000$ confirmed exoplanets and a further $>7000$ candidates\footnote{See \url{https://exoplanetarchive.ipac.caltech.edu/}}. However, not a single exomoon has been confirmed. This is in spite of a few candidates from the Kepler data \citep{Kipping2012,Teachey_2018, exomoonTTVs}, ongoing searches with the James Webb Space Telescope \citep{Christiaens2024_moondisc} and CHEOPS \citep{Ehrenreich2023}, and proposed observations with the Nancy Grace Roman Space Telescope \citep{Bachelet_exomoons}.  


Exomoon detection would be monumental for our broader understanding of planetary system formation. The most prominent theory regarding our own Moon's formation\footnote{For a thorough review on current formation theories of the Moon, see \cite{canup2021originmoon}.} is the giant impactor hypothesis \citep{GiantImpact1, GiantImpact2}, which has been more recently updated to include the effects of fast protoplanetary rotation during the collision \citep{HEmodel1, Lock_2018}. Natural satellites of gas giants are thought to have a different formation pathway, where such moons would form directly from dust within the circumplanetary disc itself \citep{diskFormation1, Batygin_2020, diskForm2}.



A planet's Hill radius, within which moons could be formed and maintained, scales linearly with the planet's semi-major axis. Therefore, searching for moons around longer-period planets would make their discovery more probable, since these moons would have more room to remain stable (though tidal instability and other secular destabilizing effects may influence a moon's long-term detectability). This is challenging as the most prolific exoplanet discovery technique--transits--is heavily biased towards short-period planets, and phase folding lightcurves to reveal exomoon signatures remains difficult \citep{Kipping_2021}. Other proposed exomoon detection methods, such as TTVs and TDVs, may also prove challenging \citep{Heller2016}.

\begin{figure}
    \centering
    \includegraphics[width=0.99\linewidth]{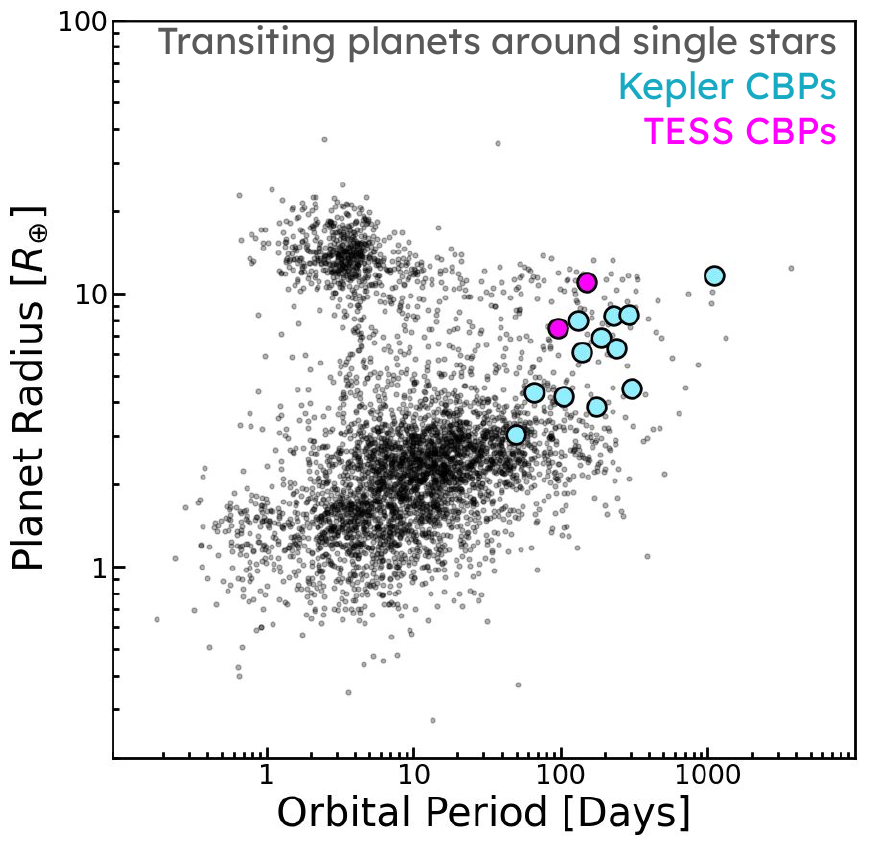}
    \caption{Radius and orbital period of all confirmed transiting planets (black dots), compared with the 12 Kepler (blue circles) and 2 TESS (pink circles) circumbinary planets.}
    \label{fig:CBP_demographics}
\end{figure}

One creative avenue is to look for moons around circumbinary planets (CBPs), as discussed in \citet{Quarles_2012}. These planets are typically found at ``long'' periods, at least relative to the rest of the transiting population (see Fig.~\ref{fig:CBP_demographics}). This is due to a combination of observational biases \citep{MartinTriaud2014,MartinTriaud2015} and planet formation restrictions \citep{Paardekooper2012,PierensNelson2013,MartinMazehFab2015,MunozLai2015}. Furthermore, a significant fraction of confirmed CBPs orbit within the habitable zone \citep{BHZ2, BHZ4, CBPHZ1, BHZ3, BHZ1}. All of the known CBPs are gas giants ($R_{\rm p}>3R_\oplus$, Fig.~\ref{fig:CBP_demographics}). This might be simply an observational bias as lightcurve photometry is more sensitive to large planets at a given orbital distance \citep{Li2016,Martin2021}, but we are nevertheless motivated to search for exomoons as they may be the best chance for CBP habitability given that gas giants are likely unable to host life.

A majority of the known CBPs orbit roughly as close as possible to the binary without being unstable \citep{MartinTriaud2014,Quarles_2018, georgakarakos2024}. This could pose challenges for the existence of a stable exomoon, as the likelihood of encountering a destabilizing mechanism dramatically increases with binary proximity \citep{martin2022runninggauntletsurvival}. \citet{Hamers2018} showed that the known CBPs could host an exomoon, as long as the moon had an orbital inclination smaller than 40$^\circ$ and was well within the planet's Hill radius. However, this study did not consider that the planet (and its moon) likely did not form in their current location. The more recently adapted theory for CBP formation is that these planets did not form \textit{in situ} \citep{Quintana_2007, Paardekooper2012}, but rather formed farther out in the disc before migrating in and stopping near a truncated inner circumbinary disc edge \citep{ArtymowiczLubow1994,PierensNelson2013}. An exomoon around a CBP would therefore represent a dynamically evolving, close to instability, four-body system. Given these conditions, a natural question arises: \textit{can circumbinary exomoons remain gravitationally bound to their host planet throughout this migration process}?


This work investigates the evolution and stability of circumbinary exomoons within the protoplanetary disc lifetime, accounting for the effects of disc-driven planetary migration and planetary oblateness. First, we will motivate key concepts that influence the dynamics of our solar systems and summarize the four archetypal outcomes (Sect.~\ref{sec:fundamental}). Next, we describe all parameter inputs used to synthesize unique binary-planet-moon populations (Sect.~\ref{sec:popsynth}). After this, we present the results of our $N$-body simulations (Sect.~\ref{sec:results}) and offer analysis alongside future implications (Sect.~\ref{sec:discussion}), before concluding (Sect.~\ref{sec:conclusion}). 

\section{Fundamental Concepts}\label{sec:fundamental}

\subsection{Type-I Migration}\label{sec:migration}

Broadly speaking, there are two regimes of disc-driven migration: Type-I and Type-II. Type-I migration comes from an exchange of angular momentum between the migrating body and a protoplanetary disc via an external torque arising from disc-planet interactions \citep{Ward1997}. To first order, the planet does not significantly alter the structure of the disc. The structure of the disc directly influences migration through an interchange between two main agents: the Lindblad and the co-rotational torques. Lindblad torque is a product of density perturbations arising from Lindblad resonances between the planet and the gas within the disc. This torque causes a planet to lose angular momentum, pushing it inwards towards the host star(s).  

\cite{lubow2010planetmigration} model the Lindblad torque in Type-I migration by: 

\begin{equation}
    T_{\rm L} = -\Sigma_{\rm d}\Omega^2 a^4 \biggl(\frac{m_{\mathrm p}}{M_{\rm A}+  M_{\rm B}}\biggr)^2 \biggl(\frac{a}{H}\biggr)^2 \ ,
\label{eqn:Lindblad}
\end{equation}
\\where $\Sigma_d$ is the density of the disc, $a$ is the semi-major axis of the migrating body, $m_{\mathrm p}$ is the mass of the migrating planet, $M_{\mathrm A}$ and $M_{\mathrm B}$ are the masses of the primary and secondary star respectively, and $H$ is the disc height. A typical disc height prescription is given as $H = ha$, where $h$ is a scaling constant, taken to be $h = 0.04$ by the authors \citep{lubow2010planetmigration}, which corresponds to the surface density of a minimum mass solar nebula (MMSN) of $\Sigma_0 = 1700$ g/cm$^2$. The disc density can be scaled by a proportionality constant, $k$, which can be varied to simulate discs of different densities, as the exact densities of protoplanetary discs are not well constrained. The Keplerian orbital frequency, $\Omega$, is given by

\begin{equation}
    \Omega = \sqrt{\frac{G(M_{\rm A} + M_{\rm B})}{a^3}} \ ,
\end{equation}
\\where $G$ is the gravitational constant. The rate at which the semi-major axis of an object decreases over time is given by the following: 

\begin{equation}
    a = a_0\rm exp\biggl(-\frac{t}{\tau_a}\biggr) \ ,
\label{eqn:adecay}
\end{equation}
\\where $a_0$ is the initial semi-major axis of the object and $\tau_a$ is the migration timescale, which for Type-I migration is calculated according to: 

\begin{equation}
    \tau_{a,I} = \frac{J}{T_{\rm L}} \ ,
\label{eqn:tau}
\end{equation}
where $T_{\rm L}$ is the Lindblad torque given by Eq.~\ref{eqn:Lindblad} and $J =m\sqrt{aG(M_{\rm A}+M_{\rm B})}$ is the object's angular momentum. A consequence of these equations is that $\tau_a\propto 1/(\Sigma_dm)$, i.e. migration is faster with more massive planets and denser discs. 

Migration also has the effect of damping the eccentricity on a much faster timescale than the semi-major axis. This timescale is found to be proportional to $\tau_{\mathrm a}$:

\begin{equation}
    \tau_{e} = \frac{\tau_{ a}}{\mathrm{C}} ,
\label{eqn:tau_e}
\end{equation}
\\where $\mathrm{C} = 10$ is a constant found in \cite{Kley2004} and used in \cite{martin2022runninggauntletsurvival}.

Whilst the Lindblad torque acts to drive a planet or moon inward, the co-rotational (or horseshoe) torque is capable of achieving the opposite effect. As the disc material nearest to the planet librates with the co-orbital horseshoe region, the resulting torque pushes the planet outwards. A key difference with the Lindblad torque is that the Lindblad torque is proportional to the \textit{disc density}, whereas the corotational torque is proportional to the \textit{disc density gradient}, i.e. $T_{\mathrm{co}}\propto \mathrm{d}\Sigma_d/dr$. In the case of binaries, there is a sharp drop-off in disc density at the truncation radius at $\sim 3a_{\mathrm{bin}}$, arising from tidal torques from the binary (see Sect.~\ref{sec:disctruncation}). There, the co-rotational torque becomes large enough to match the Lindblad torque and cease the planet's migration Sect.~\ref{sec:migration_known_CBPs}).


\subsection{Type-II Migration}\label{sec:t2migration}

If a planet is massive enough, local dust and gas will accrete onto the planetary surface, opening a `gap' in the protoplanetary disc. The migration mechanism laid out in Sect.~\ref{sec:migration} no longer applies, as there is no longer a local disc density to cause an inward torque. Nevertheless, massive planets are still thought to migrate inward. This leads to the idea behind Type-II migration, where a massive planet (roughly the mass of Saturn, or 95$\rm M_{\oplus}$) will be forced inward due to interactions with gas that is also undergoing inward migration, but on a much longer timescale than $\tau_{\rm a, I}$ \citep{Lin_1986}. 

In this work we follow the model used in \citet{Ida_2018}, which shows that Type-II migration timescale is related to the calculated Type-I migration timescale by a scaling factor: 

\begin{equation}
    \tau_{\rm a, II} \simeq \frac{\Sigma}{\Sigma_{\mathrm{min}}} \tau_{\rm a, I} \ ,
\label{eqn:type2migration}
\end{equation}
where $\Sigma$ is the unperturbed surface density of the disc, $\Sigma_{\rm{min}}$ is the surface density at the bottom of the gap. 
This scaling factor is given by

\begin{equation}
    \frac{\Sigma}{\Sigma_{\rm {min}}} \simeq (1 + 0.04K),
\end{equation}
where

\begin{equation}
    K = \biggl(\frac{m_{\rm p}}{M_{\rm A} + M_{\rm B}}\biggr)^2 \biggl(\frac{H}{\rm a}\biggr)^{-5} \alpha_{\rm{vis}}^{-1} \ .
\end{equation}
In this expression, $\alpha_{\rm{vis}}$ is the turbulent strength of the disc, and has a value of $3 \times 10^{-4}$.

In Type-I migration, the outwards co-rotational torque provides a natural stopping mechanism. Type-II migration, in comparison, does not have a comparable stopping mechanism.

\subsection{Stability Limit \& Disc Truncation}\label{sec:disctruncation}

Binary stars exert gravitational torques that clear out a central cavity in the circumbinary disc, typically extending to several times the binary separation. This truncation radius roughly scales with the `stability limit' around the binary, given by \citet{stabLimCalculation} as:
\begin{equation}
\begin{split}
    \frac{a_{\rm s}}{a_{\rm bin}} = 
    1.60 + 5.10e_{\rm bin} - 2.22e_{\rm bin}^2 + 4.12\mu  \\
    - 4.27e_{\rm bin}\mu - 5.09\mu^2 + 4.61e_{\rm bin}^2\mu^2 \ ,
    \end{split}
    \label{eqn:stabLimitCalc}
\end{equation}
in which $a_{\rm s}$ is the stability limit measured from the barycenter, $e_{\rm bin}$ is the eccentricity of the binary, and $\mu$ is the mass fraction of the binary, given by $\mu = M_{\rm A} / (M_{\rm A} + M_{\rm B})$. For reference, a binary with $e_{\rm bin} = 0$ and $\mu = 0.5$ yields $a_{\rm s} = 2.39 a_{\rm bin}$\footnote{This formula found in \citet{stabLimCalculation} includes error bars. As discussed later in this section, objects in our simulations are forced to stop at 1.5 times the calculated limit, which is well beyond the upper bound of these error bars. Therefore, we omit them here.}. 

Although the idea of a stability limit is useful in helping to conceptualize possible orbital configurations within a circumbinary system, it is not the only factor that determines a planet's orbital stability. 
Other factors, such as the planet's eccentricity, orientation, phase of the orbits, and mean motion resonances influence the overall stability \citep{Mudryk_2006,DoolinBlundell2011,Kratter_2014, Smullen_2016}. To conservatively account for these complexities, we chose to set the inner disc edge (or the disc truncation radius) of the migrating object at 1.5 times the calculated stability limit, as we are only interested in scenarios in which the host planet's orbit remains stable. It was found through trial and error that a scaling factor of 1.5 allows for planetary orbital stability during its migration and any subsequent evolution despite these various destabilizing effects. Various works have sought to update this model for the stability limit, notably for a larger parameter space \citep{Quarles_2018} and using machine learning techniques \citep{Lam_2018}. In this work, we focus on the model outlined in Eq.~\ref{eqn:stabLimitCalc}, as our systems migrate to $1.5a_{\rm s}$, and thus the outcome would not be sensitive to such a change.


\subsection{Migration of Circumbinary Planets}\label{sec:migration_known_CBPs}

Of the confirmed transiting CBPs, the majority are near the stability limit around their binary. This limit roughly coincides with the inner edge of the protoplanetary disc, where the binary carved out an inner ``hole'', calculated in \citet{ArtymowiczLubow1994} and inferred observationally in \citet{CircumbinaryDisks}. CBPs likely did not form \textit{in situ} because the protoplanetary disc would have been too turbulent close to the inner edge of the disc \citep{Paardekooper2012}. They instead probably formed farther out in the disc before migrating inwards towards the disc edge, parking near their observed locations \citep{CircumMigrationPierens2008,PierensNelson2013,planet_parking_Penzlin_2021,martin2022runninggauntletsurvival}. 

With respect to Type-I and Type-II migration, \citet{CircumMigrationPierens2008} put forward an early argument that a CBP following Type-II migration would not have a co-rotation torque to brake the inwards migration at the disc edge, because the horseshoe region of the disc has been cleared out. Such planets would migrate too close to the binary and be ejected. Therefore, any planets found near the inner edge of the disc must have followed Type-I migration. Indeed, of the known planets, the ones closest to the disc edge are all less massive than Jupiter, although their masses are typically poorly constrained. Whilst we acknowledge uncertainty concerning the prevalence of Type-II migration for CBPs, the migration of planets that are more massive than Saturn in our simulations is modeled under the Type-II regime and is enforced to end their migration at the disc's truncation radius, to match observational trends.

In our study, we implement migration using \textsc{ReboundX} \citep{reboundX}, which contains add-ons to the \textsc{Rebound} N-body integrator \citep{rebound,Reboundintegrator}. Our implementation is similar to \citet{martin2022runninggauntletsurvival} and \citet{fitzmaurice2022}, except for two simplifications. First, we use the newly-released \textsc{inner disc edge} functionality of \textsc{ReboundX} to halt migration. Second, for a given circumbinary system we use a constant migration rate, calculated from Eq.~\ref{eqn:Lindblad} for Type-I migration ($M_{\rm p}<95M_{\oplus}$) and Eq.~\ref{eqn:tau} for Type-II migration ($M_{\rm p}>95M_{\oplus}$), evaluated at 1 AU. Ideally, the torque would change with the migration of the planet. However, this was found to be computationally expensive for our large suite of million-year simulations. Given that the migration timescale will change by orders of magnitude between simulations of different disc densities and planet masses, we found this simplification to be acceptable.


\subsection{The Hill Sphere \& Roche Radius}\label{sec:Hillsphere}

A planet's Hill sphere is defined by the region around the planet where its gravitational influence on a test body largely dominates over the competing gravitational force from the planet's host star(s). The Hill radius, $r_{\rm Hill}$, which represents the approximate boundary of this sphere, is given by:

\begin{equation}
    r_{\rm Hill} = a_{\rm p} \left[\frac{m_{\rm p}}{3(M_{\rm A}+M_{\rm B})} \right]^\frac{1}{3}\ .
\label{eqn:Hillradius}
\end{equation}
We model the Hill radius of a body independent of eccentricity, as we assume that the planets formed beyond several AU in the disc have circular orbits. Also, the eccentricity damping timescale, given in Eq. \ref{eqn:tau_e}, is a factor of 10 smaller than the migration rate, meaning that a planet's orbit will quickly become circular if not already. 

In the context of our system, the Hill radius of the planet decreases as it migrates inwards. Therefore, the survivability of the moon is directly influenced by how much the Hill radius shrinks. It naturally follows that there should exist a maximum distance a satellite can orbit its host planet without being ejected during the migration process when only considering Hill radius effects. Previous studies have shown that it is \textit{possible} for a moon in a single star system to remain in a stable orbit around its host planet with its orbital radius as large as 0.48 $r_{\rm Hill}$ \citep{Domingos2006, Rosario-Franco2020}. Therefore, when we generate system parameters for subsequent analysis, we do not want to populate moons that are unstable upon creation. For this reason, we use a distance of 48$\%$ of the host planet's initial Hill radius as the upper bound to where moons can be populated.

Due to planetary migration, the stability of a moon is also sensitive to the orbit of the host binary, although in an implicit way. As previously discussed, a planet halts its migration as it approaches the truncation radius of the disc. The truncation radius scales roughly with the binary's stability limit, which in turn depends on the binary's semi-major axis, mass fraction, and eccentricity (see Eq.~\ref{eqn:stabLimitCalc}). Therefore, planets in tight binary systems ($\sim10$ day binary period) are allowed to migrate closer, resulting in smaller Hill radii at the end of their migration, in comparison to those hosted by longer period binaries.

Given that a planet is capable of maintaining moons up to 48\% of its Hill radius, allowing a planet to migrate closer to the binary decreases the maximum value of $a_{\rm moon}$ at which the moon can remain stable. Therefore, the smaller the stability limit of the binary, the higher the likelihood of a moon being ejected when considering planets that migrate completely to the inner disc edge. If the moon escapes the planet's Hill sphere and is ejected, it will undergo standard Type-I migration.

Similar to how the Hill radius acts as an ``outer limit'' for a satellite's orbital stability, the Roche radius (or Roche limit) can be seen as the corresponding ``inner limit''. The Roche radius in this context is defined as the orbital region at which the self-gravitation holding a satellite together is overcome by the gravitational potential well of the planet, in turn ripping apart the satellite.

The Roche radius, $R_{\mathrm{Roche}}$, is given by the following: 

\begin{equation}
    R_{\mathrm{Roche}} = \biggl(\frac{2\rho_\mathrm{p}}{\rho_\mathrm{s}}\biggr)^{\frac{1}{3}} R_{\mathrm{p}}
\label{eqn:Roche}
\end{equation}
where $\rho_\mathrm{p}$ and $R_\mathrm{p}$ are the density/radius of the planet, and $\rho_\mathrm{s}$ is the density of the satellite. Note that this equation is a function of both the planet and satellite densities, as moons with larger relative densities will be more difficult to break apart, and as a result, the Roche radius would lie closer to the center of mass of the planet. If a moon is more than twice as dense as its host planet, or $\rho_\mathrm{s} > 2\rho_\mathrm{p}$, the Roche radius would lie within the planet's radius, as could be the case for a gas giant with a rocky (dense) moon.



\subsection{Planetary Oblateness}\label{sec:oblateness}
Planetary oblateness can cause the moon's orbit to undergo apsidal precession about the host planet, creating a source of instability on timescales much less than the disc lifetime \citep{Spalding_2016}. Particularly, it is shown that moons orbiting closer than 10 $R_{\rm p}$ of a migrating Jupiter-like planet about a single star would be ejected due to excessive eccentricity growth due to `evection resonance'. This resonance occurs when the rate of the moon's apsidal precession becomes equal to the host planet's orbital period. As a result, the moon's orbital eccentricity about the planet rapidly grows, leading to eventual ejection.


We adopt the \textsc{gravitational harmonics} package in \textsc{ReboundX} to model the precession caused by an oblate host planet, implemented by \citet{Tamayo2020}. We take the zonal harmonic coefficient, $J_2$, to be 0.02 for $M_{\rm p} \ge 95M_{\oplus}$ (i.e.- 1 Saturn mass) as performed in \citet{Spalding_2016}, and $1.083 \times 10^{-3}$ for planets below this threshold \citep[see Table A.4 in][]{Murray1999}. We set the equatorial radius for each planet as the planetary radius sampled from the known exoplanet distribution (see Sect. \ref{sec:popsynth}). Lastly, we assume the planet's spin axis is perpendicular to the orbital plane.


Over the Myr timescales of a typical protoplanetary disc, the effects of tides or spin-induced migration are negligible, as these processes occur on much longer (Gyr) timescales. Therefore in our simulation, if the moon is bound to a planet, it does not experience migration relative to that planet\footnote{However, all stable moons at the end of our simulations may be subject to these effects beyond the disc dissipation timescale (see Sect.~\ref{sec:caveats})} (except for capture into the evection resonance). Like previous studies by \citet{Namouni2010,Spalding_2016,Pu2025,Bolmont2025}, we consider disc-driven migration of the moon relative to the planet to be negligible and ignore it. We provide a rough calculation justifying this approximation in the Appendix.


\section{POPULATION SYNTHESIS}\label{sec:popsynth}




The four-body orbital evolution is chaotic and therefore highly sensitive to initial conditions. The demographics of circumbinary planets, particularly at their birth, are also very uncertain. To account for this, we utilize a broad population synthesis study. Here, we generate initial parameters for the Binary-Planet-Moon system and the protoplanetary disc. We do this via Monte Carlo sampling of reasonable values\footnote{For an updated list of properties for confirmed CBPs and the corresponding star systems, see Table 5 in \citet{georgakarakos2024}.} in the following way. The primary star mass was sampled uniformly from the current distribution of binaries that host CBPs ($0.69$ to $1.53 \ M_\odot$). To calculate the secondary star mass, we sampled the stellar mass ratio, $q=M_{\mathrm B}/M_{\mathrm A}$, uniformly between 0.1 to 1. If this produced a mass below the stellar mass threshold ($0.08M_\odot$), we substituted $0.19 M_\odot$, which is the lowest mass star known to host a CBP. We sample the binary period from the Kepler binary period distribution from the smallest CBP hosting binary to the largest (7 to 41 days). The binary eccentricity is drawn from a Rayleigh distribution centered on the mean value of all known binaries if P$_{\rm bin} \ge 12$ days, and 0 otherwise \citep{Raghavan_2010}. 

We sample the planet's mass by first drawing uniformly from the radius distribution for all currently known exoplanets. We then calculate a planetary mass by applying the mass-radius relationship found in \cite{Chen_2016}. The planet's starting semi-major axis is uniformly sampled from 1 to 5 AU about the barycenter. If the 4:1 period resonance with the calculated stability limit is beyond 1 AU, this was then set as the lower limit, as destabilizing resonances could prevent planet/moon formation in this region \citep{martin2022runninggauntletsurvival}. 

The moon's semi-major axis is sampled uniformly between the two boundaries described in Sect~\ref{sec:Hillsphere}: the Roche limit and 0.48 of the initial Hill radius. The values for the Roche limit varied in our sample between 0.0002 and 0.004 of the initial Hill radius. True anomaly and argument of periapsis were sampled uniformly from 0 to 360 degrees for both the planet and the moon. In the case that a moon escapes the planet's Hill radius and migrates on its own following Type-I migration, its migration timescale is calculated assuming that the moon's mass is similar to Ganymede's. Given that ice giants are thought to be the most prominent planet type beyond the snow line \citep{Suzuki_2016}, it is not unreasonable to assume that these giants can host Ganymede-sized moons. Since a typical moon's mass is at least 2-3 orders of magnitude smaller than a planet's, we set it as a massless particle in the $N$-body simulations to reduce the computational time.

\begin{figure}
    \centering
    \includegraphics[width=0.99\linewidth]{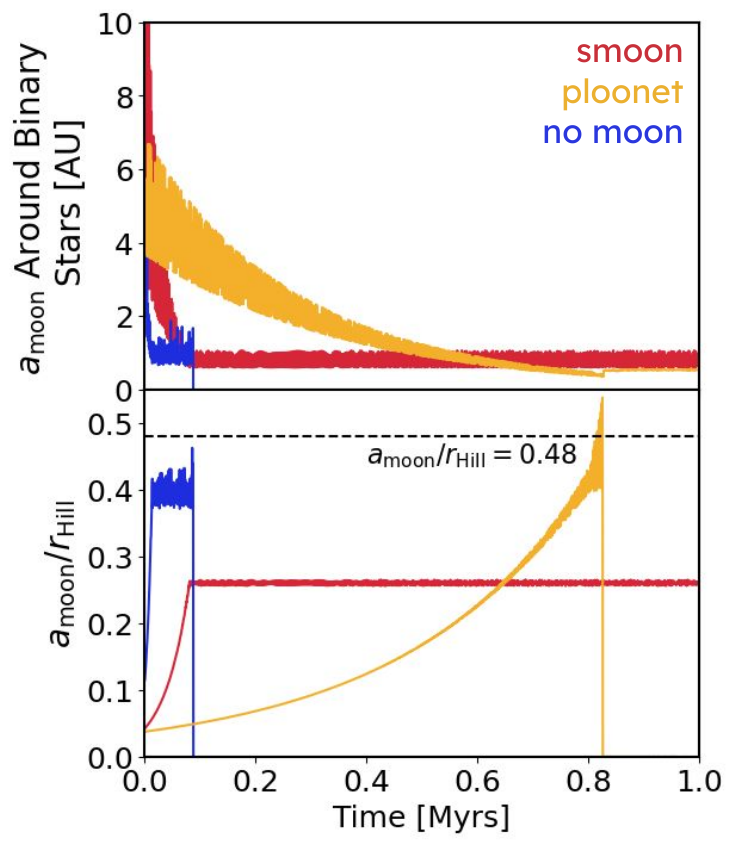}
    \caption{{\bf Top:} Semi-major axis of the moon with respect to the binary for three archetypal systems: a saved moon (``smoon'', red), a moon-turned-planet (``ploonet'', yellow), and a completely ejected moon (``no moon'', blue). The three samples shown display ideal conditions and are not used in our final dataset, hence starting beyond 5 AU. Since the moon orbits a planet, its osculating semi-major axis calculated relative to the binary stars oscillates around its host planet's orbital radius, as can be seen as the thick parts in the plot. The ploonet example shows a moon that is initially orbiting the planet but eventually leaves the planet's Hill sphere after 0.8 Myr and starts to orbit the binary as a CBP. \\{\bf Bottom:} Time evolution of $a_{\mathrm{moon}}/r_{\mathrm{Hill}}$ for the same three archetypal systems. For the retained moon (red), $a_{\mathrm{moon}}/r_{\mathrm{Hill}}$ initially increases due to $r_{\mathrm{Hill}}$ decreasing as the planet migrates inwards. At roughly 0.1 Myr, the planet reaches the disc truncation radius, so the $a_{\mathrm{moon}}/r_{\mathrm{Hill}}$ ratio plateaus at 0.26. For the ploonet (yellow) and ejected moon (blue), $a_{\mathrm{m}}$ becomes undefined about the planet once the moon moves beyond the Hill radius.}
    \label{fig:archetypes}
\end{figure}

\begin{figure}
    \centering
    \includegraphics[width=0.99\linewidth]{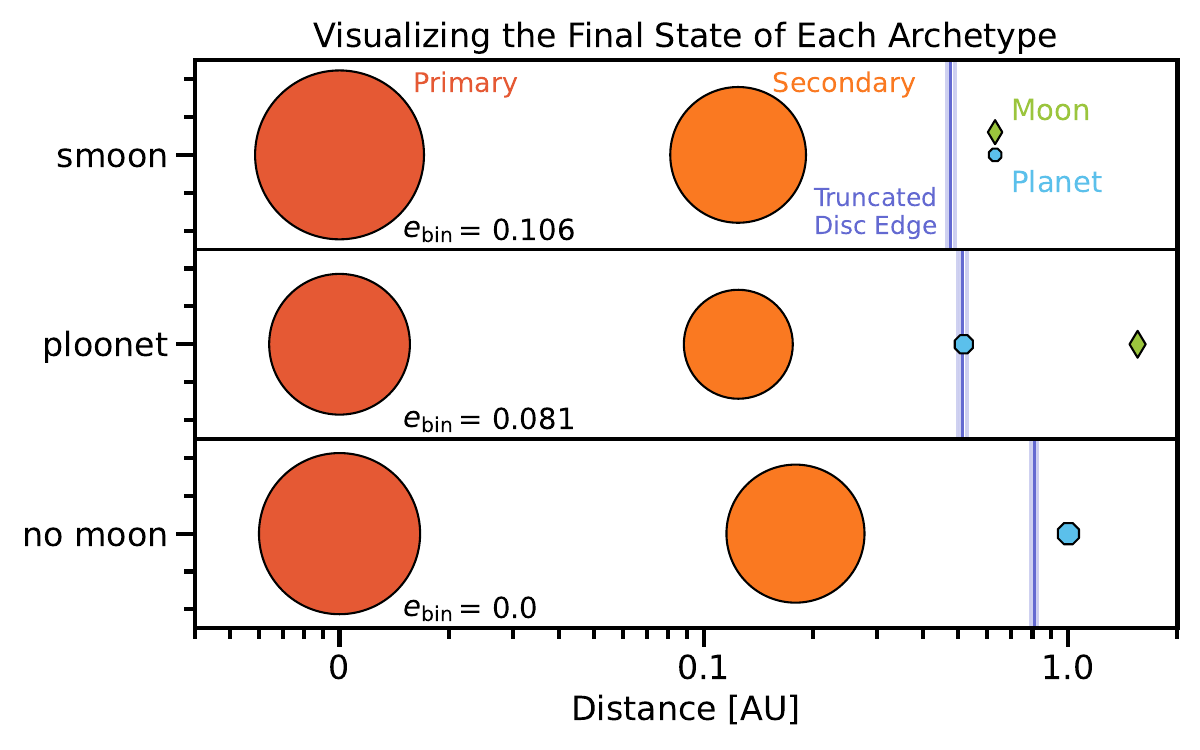}
    \caption{Three example simulations demonstrating a saved moon, ``smoon'' (top), a moon-turned-planet, ``ploonet'' (middle) and an ejected moon, ``no moon'' (bottom). The distance scale is logarithmic. The depicted sizes of the stars and the planet are scaled by the body's mass. The stability limit (and corresponding error bars) pictured by the vertical purple line is 1.5 times the limit calculated via the criteria in \cite{stabLimCalculation}, with the error bars taken from the same paper.}
    \label{fig:visualizations}
\end{figure}

The stability limit of the binary is calculated from Eq.~\ref{eqn:stabLimitCalc}. We multiply this by 1.5 to calculate the disc truncation radius (see Sect.~\ref{sec:migration}). The migration rate of the planet is using Eq.~\ref{eqn:tau} for planets with $M_{\rm p}<95M_{\oplus}$ and Eq.~\ref{eqn:type2migration} for more massive planets. Given a migration rate, the damping of eccentricity is found via Eq.~\ref{eqn:tau_e}.

From Eq.~\ref{eqn:Lindblad}, the Lindblad torque is proportional to the disc surface density. The surface density is scaled from $\Sigma_{\rm d} = 1700 k \ \rm{g/cm}^2$, with the scaling factor $k$ sampled from a log-uniform distribution from -2 to 2, to cover a wide range of disc densities. Based on \citet{diskLifetime2, binarydisklifetime}, we use a constant disc lifetime of 1 Myr, after which the disc is assumed to dissipate and the simulation is ended. 

We use the \textsc{Rebound} package to perform $N$-body integration of our binary-planet-moon systems using \textsc{IAS15} \citep{Reboundintegrator}. We measure the orbital parameters of the planet and the moon every 100 years of orbital evolution, but decrease this value to 2 hours for more accurate tracking of planet-moon close encounters. We employ the \textsc{Reboundx} library to incorporate the effects of planetary migration \citep{mofbasedon, modifyorbitforces}, the inner disc edge \citep{IDEbasedon, IDEimplementation}, and planetary oblateness \citep{Tamayo2020}.

\section{RESULTS}\label{sec:results}

We utilize the \textsc{Rebound} $N$-body integrator to simulate the orbital evolution of 2000 binary, circumbinary planet, and exomoon systems over 1 Myr of disc evolution. In Sect.~\ref{sec:archetypes} we outline the four broad categories of outcomes. In Sects.~\ref{sec:results_moonorbit} and \ref{sec:results_migtimescale} we discuss the results as a function of the moon orbit and the migration timescale, respectively. The occurrence rate of each outcome is cataloged in Table \ref{tbl:results}.

\begin{figure*}
    \centering
    \includegraphics[width=0.99\textwidth]{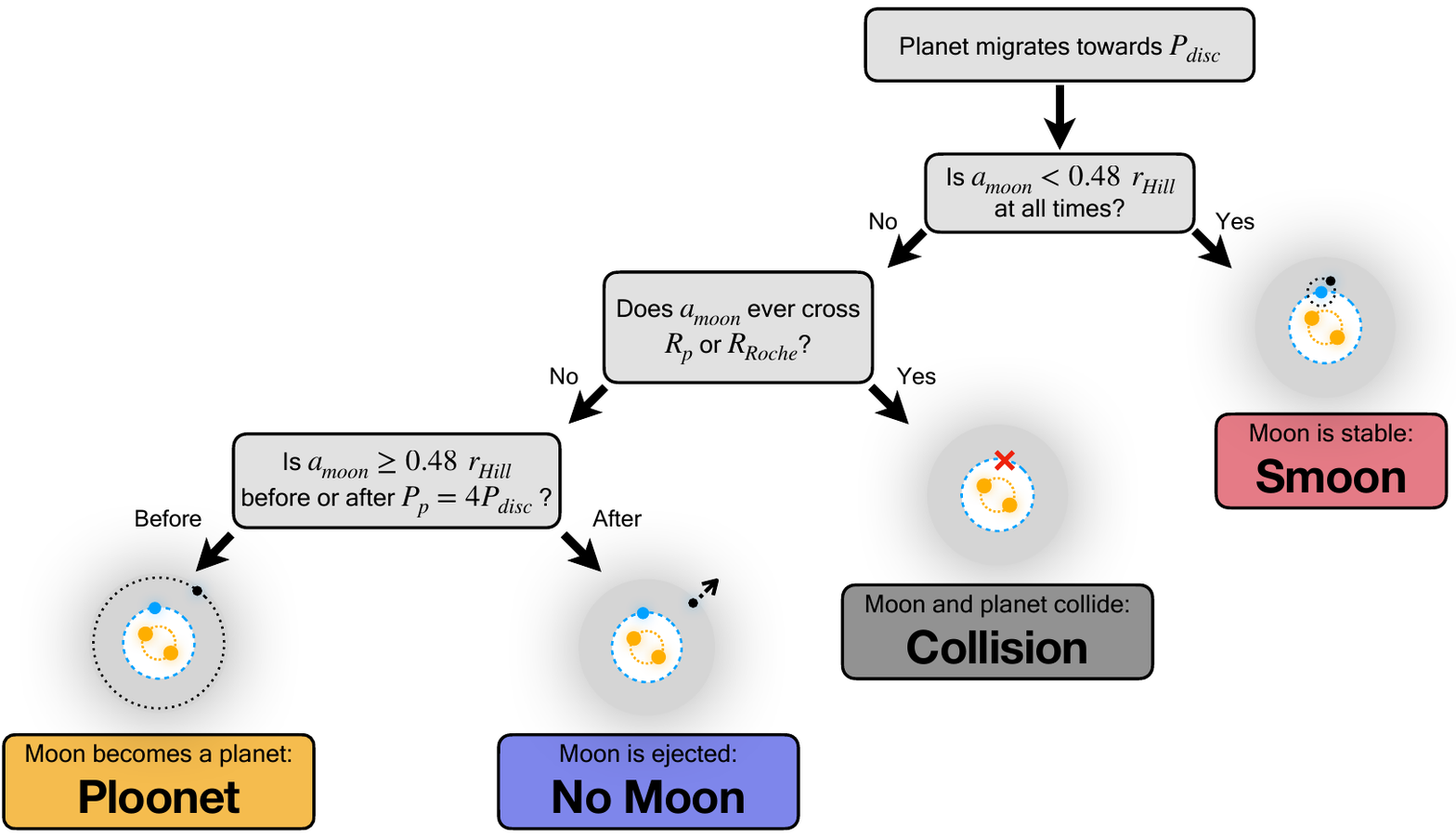}
    \caption{Flowchart showing the four expected outcomes of a moon orbiting a circumbinary planet that migrates in towards the inner edge of the circumbinary disc (denoted by an orbital period of $P_{\rm disc}$). If the planet does not migrate completely to the disc edge, then the moon is more likely to be saved (a ``smoon''), as the first bifurcation criterion is likely to be met.}
    \label{fig:flowchart3}
\end{figure*}

\begin{table*}
    \centering
    \caption{The occurrence rate of each outcome for our simulations. Each of the percentages is given regarding the total population of systems. From this, we can see that the ploonet is the most likely of the outcomes. However, planets that only partially migrate are more likely to retain their moon than those that migrate to within $5\%$ of the calculated stability limit. Orbital elements of the planet-moon system are not tracked post-collision, so there are no migration statistics for this case. In total, 33$\%$ of planets migrated to within $5\%$ of the calculated stability limit. Of the 686 `partially migrated' planets, 321 (47$\%$) migrated to within the 10:1 period resonance with the binary, where dynamic perturbations from the binary begin to affect orbital stability.}
    \begin{tabular}{cccccc}
         \hline
         \textbf{Population}\quad & \textbf{Total}\quad & \textbf{Ploonets (\%)} \quad& \textbf{Smoons (\%)} \quad& \textbf{No Moon (\%)} \quad& \textbf{Collision (\%)}\\
         \hline
         \hline
         All Systems & 2000 & 758 (38) & 578 (29) & 14 (1) & 650 (32) \\
         Fully Migrated & 664 & 466 (23) & 186 (9) & 12 (1) & - \\
         Partially Migrated & 686 & 292 (15) & 392 (20) & 2 ($<1$) & - \\
    \label{tbl:results}
    \end{tabular}
\end {table*}

\subsection{System Archetypes}\label{sec:archetypes}

$N$-body simulations yield four main outcomes for the binary-planet-moon system. 

\begin{enumerate}
    \item {\bf Smoon:} If the moon is well within the Hill radius $(a_{\mathrm {moon}}/R_{\mathrm {Hill}}  < 0.48)$ of the planet throughout the entirety of the migration, the planet will retain its moon, leading to a single planet system with a moon. This system will be referred to as the successful moon, or the `smoon' scenario for short. See the red curves in Fig.~\ref{fig:archetypes} and the top example in Fig.~\ref{fig:visualizations}.
    \item {\bf Ploonet:} During the planet's migration, its Hill radius can decrease such that the moon's orbit becomes unstable when $a_{\mathrm {moon}}/R_{\mathrm {Hill}} \approx 0.48$. As a result, the moon escapes the planet's Hill sphere. The body, formerly known as a moon, is now a planet in its own right.\footnote{We adopted the term `ploonet' for the moon-turned-planets, following \cite{ploonet2_Montes_2017} and \cite{ploonet_Sucerquia_2019}.} The ploonet interacts with the disc and undergoes Type-I migration. Its migration speed is much slower than that of the original planet, as $\tau_a\propto 1/m$. Since the migration is so slow, many of the ploonets are on a wide orbit at the end of the simulation, far from the binary and planet (See Fig.~\ref{fig:ploonetHist}). See the yellow curves in Fig.~\ref{fig:archetypes} and the middle example in Fig.~\ref{fig:visualizations}. In some sporadic cases, the moon can migrate in and get locked in an exterior mean motion resonance, similar to what was seen in \citet{fitzmaurice2022}.
    \item {\bf No moon:} The moon is entirely ejected from the system, leaving behind a moonless circumbinary planet. This scenario typically occurs when the moon escapes the planet relatively close to the disc's truncation radius (approximately $P_{\rm p}\leq4P_{\rm disc}$ or within). 
    In this case, the moon cannot turn into a stable ploonet due to destabilizing and overlapping resonances relative to both the planet and the binary \citep{MudrykandWu2006, Sutherland_2019}. See the blue curves in Fig.~\ref{fig:archetypes}  and the bottom example in Fig.~\ref{fig:visualizations}.
    \item {\bf Collision:} The moon collides with the host planet, or crosses the Roche radius (see Sect.~\ref{sec:Hillsphere}, Eqn.~\ref{eqn:Roche}), and is ripped apart. Whilst the latter scenario does not involve a \textit{physical} collision, it has the same effect of destroying the moon, and so both scenarios are categorized in the same way. A collision can occur during the ejection process or later on in disc evolution after temporarily becoming a ploonet (see Table~\ref{tbl:results} and gray points in Fig.~\ref{fig:ganyMoon1}). The occurrence rate of this result is slightly elevated but consistent with planet-moon collisions tracked in similar studies (see Table 3 in \citealt{Trani_2020}).
\end{enumerate}

For a given starting planet and moon, and an assumption that the planet has time to migrate to the disc's edge, one can typically predict the outcome for the moon based on a flowchart shown in Fig.~\ref{fig:flowchart3}. Some planets will not migrate to the disc edge within the 1 Myr simulation time. This is often the case for low-mass planets in low-density discs and gas giants which undergo Type-II migration (see Table~\ref{tbl:results} for occurrence rates; small/large red dots in Fig.~\ref{fig:ganyMoon1}). Their slow migration rates lead such planets to end their migration farther away from the binary within the disc's limited lifetime. As a result, they end up with larger Hill radii and thus would be more likely to retain their moons compared to their fully-migrated counterparts that end their migration near the disc's truncation radius.

\subsection{Dependency on Moon Orbit}\label{sec:results_moonorbit}



\begin{figure}
    \centering
    \includegraphics[width=0.99\linewidth]{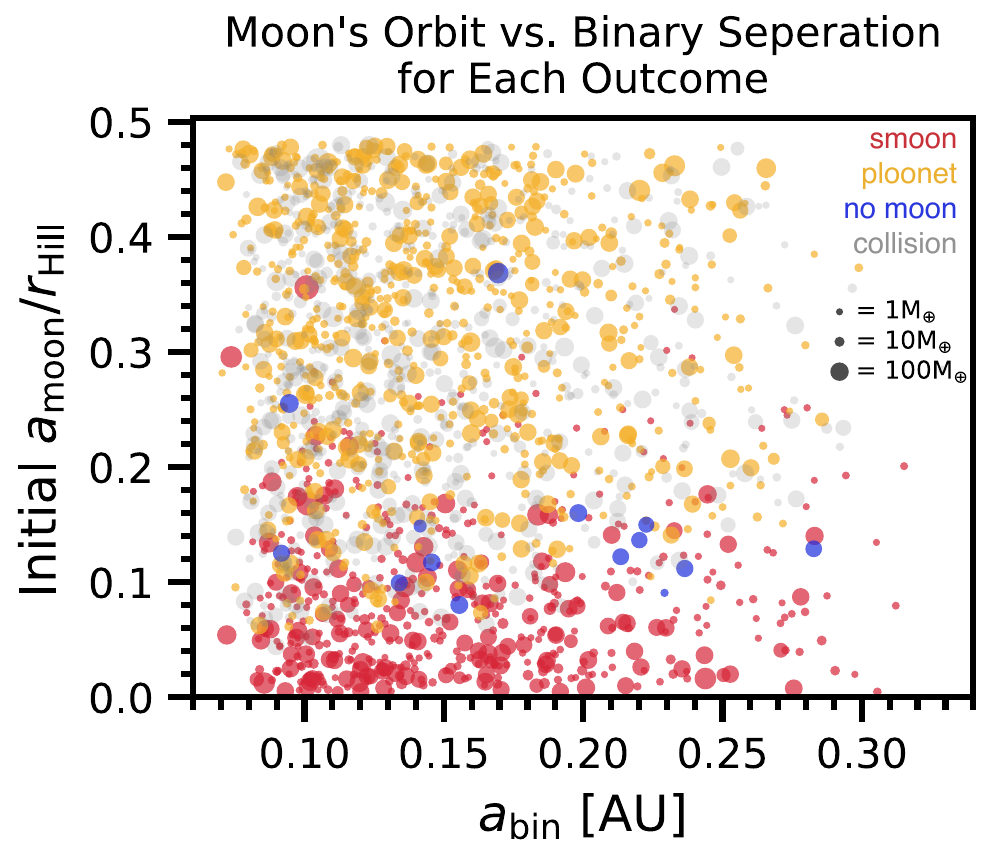}
    \caption{Results from $N$-body simulations showing how the initial moon-planet and planet-binary separation impacts the outcome. The smoon, ploonet, no moon, and collision archetypes are labeled via red, yellow, blue, and gray dots, respectively. The dot size corresponds to that planet's mass on a log scale, as depicted. As predicted, moons that are initially very close to their host planet are retained, whereas moons initially far from their host planet easily become unstable. The binary separation also plays a critical role, as a wider binary makes it more likely for a planet to retain its moon. Less massive planets did not migrate as far and were able to retain their moons, as seen by small red dots with a large $a_{\mathrm {moon}}$ to $ r_{\mathrm {Hill}}$ ratio. The collisional outcome populates all regions of parameter space in which moons are ejected from their host planet.}
    \label{fig:ganyMoon1}
\end{figure}


The distribution of moon starting semi-major axis scaled by the planet's initial Hill radius versus the binary stars' orbital radius sampled for each planetary population, color-coded by the moon's final outcome, is shown in Fig.~\ref{fig:ganyMoon1}. General trends of the outcome are as follows. Moons with a small initial $a_{\rm {moon}}$/$r_{\rm {Hill}}$ ratio are more likely to remain a satellite of their host planet after a 1 Myr phase of migration. This is the case for moons in relatively tight orbits around their host planet, since the ratio between their orbital radius to the planet's Hill radius never exceeds the threshold $ a_\mathrm{moon}/r_{\rm Hill}\approx0.48$. This is consistent with results in \citet{RabagoSteffen2018} (see their Fig. 2), who show that perturbed moons have a higher survival probability the closer they orbit within a planet's gravitational potential well.

The initial $a_{\mathrm {moon}}$/$r_{\rm {Hill}}$ ratio that allows the moon to end up as a smoon also increases from 0.5 to 1.5 as the binary stars' separation, $a_{\rm bin}$ increases from 0.1 to 0.3 AU. This is because the planets orbiting wide binaries end their migration at a larger distance from the binary stars. Thus, their final Hill radius remains large, allowing moons in a wider orbit to remain in a stable orbit as the planets' satellites.

The majority of planets in our simulations do not retain their moons. As mentioned in Sect.~\ref{sec:archetypes}, moons that leave the Hill sphere of the planet close to the end of the planet's migration are likely to end up in an unstable orbit and will eventually be ejected. A few cases were observed in our simulations (the ``no moon" archetypes). Most of these systems occupy the intermediate region between the smoons and the ploonets in Fig.~\ref{fig:ganyMoon1}. 

It is also possible that moons collide into the host planet upon ejection. To test this, we track the orbital elements of unstable moons on a much shorter timestep during/after the ejection (hours rather than years). From this analysis, we found that $32\%$ of our simulations involve planet-moon collisions. Moons are likely to collide with their host planet during the ejection process, though certain cases involve a moon being ejected on an interior orbit, meaning the subsequent planet-ploonet collision occurs later on in the disc evolution. 

\subsection{Dependency on Migration Timescale}\label{sec:results_migtimescale}

\begin{figure}
    \centering
    \includegraphics[width=0.99\linewidth]{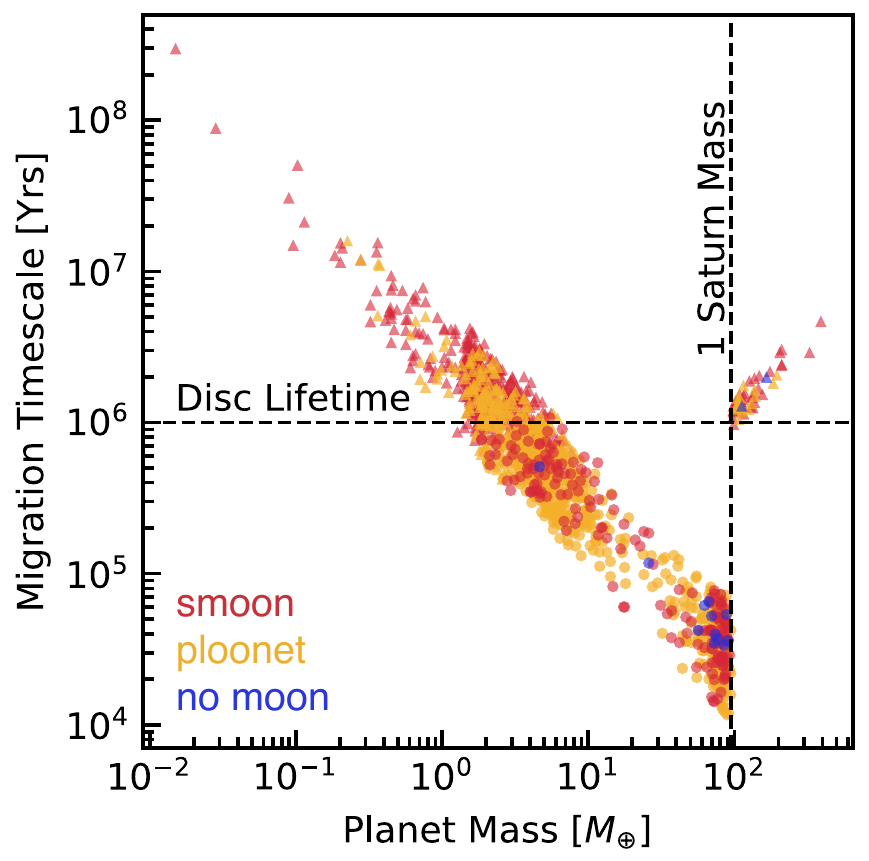}
    \caption{A planet's ability to migrate to within $5\%$ of the truncated disc edge is largely dependent on its mass. The triangles correspond to planets that did not migrate to within 5$\%$ of the disc truncation radius ($1.5a_{\mathrm s}$), whilst circles correspond to planets that migrated within this limit. The red, yellow, and blue markers correspond with the smoon, ploonet, and no moon scenarios, respectively. The no moon scenario is more common for high-mass planets undergoing Type-I migration, as the large Hill radii of these planets allow them to retain their moons until they are close to the binary. No planet that underwent Type-II migration (any planet above a Saturn mass) reached within 5$\%$ of the disc truncation radius.}
    \label{fig:migTimescale}
\end{figure}

A portion of the planets in our simulation undergo migration on a timescale close to or longer than the disc lifetime and thus never reach the disc truncation radius, as depicted by triangle markers in Fig.~\ref{fig:migTimescale}. None of the planets undergoing Type-II migration, indicated by a Saturn mass, reach the inner edge of the disc within the 1 Myr disc lifetime. There is mixing for systems with migration timescales slightly less than 1 Myr due to a uniform sampling of initial planetary orbital distances and disc densities.

Ploonets have the highest occurrence rate of all outcomes. Moons that leave the planet's Hill sphere whilst still far from the binary are the most likely to remain stable. The distribution of the final orbital period of the surviving ploonets is shown in Fig.~\ref{fig:ploonetHist}. Most of them are in multi-year orbital periods. This is because the moons in our sample are approximately 2 to 5 orders of magnitude less massive than the planet, with a migration timescale on the order of 10 to 100 Myr. These migration timescales are much smaller than the 1 Myr lifetime of the disc, so ploonets hardly migrate from where they were ejected. If the ploonet migrated faster (or for longer), it could be trapped in an exterior mean motion resonance with the planet, as seen in \citet{fitzmaurice2022}. However, given the slow migration of a moon-mass object, this was a rare event.

\section{DISCUSSION}\label{sec:discussion}

\subsection{Multi-Planet Circumbinary Systems}

\begin{figure}
    \centering
    \includegraphics[width=0.99\linewidth]{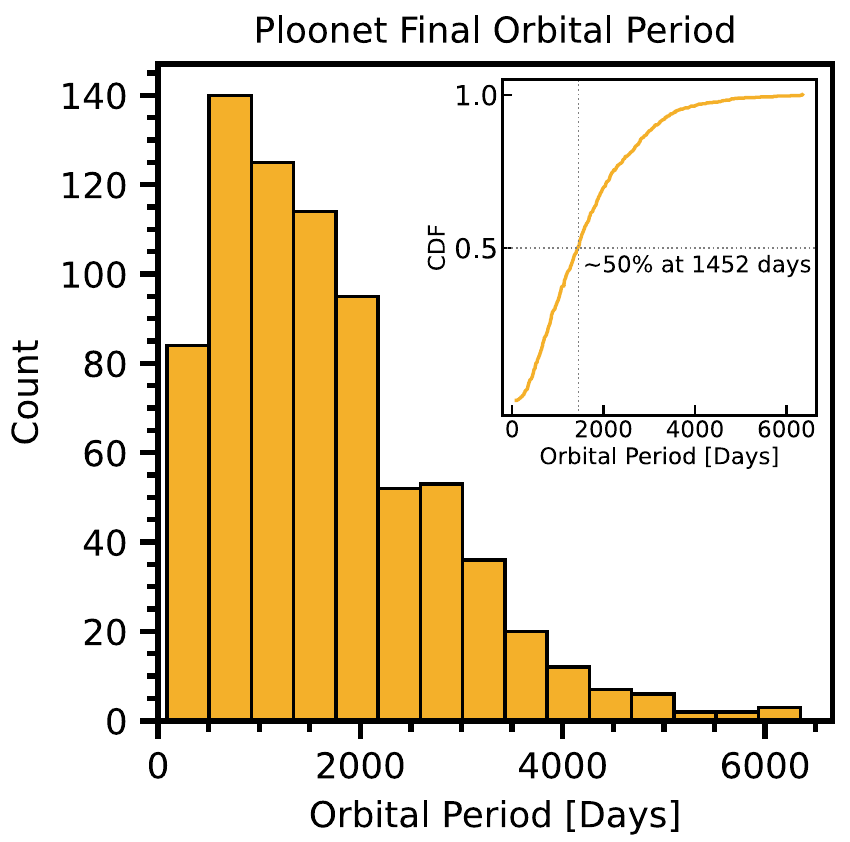}
    \caption{Orbital period distribution of all stable ploonets at the end of 1 Myr evolution. 7 systems had periods above 7500 days, which are not included in this distribution. Over half of the ploonets have periods above 1400 days, much larger than the current population of known CBPs. It is unlikely these ploonets would be detected via Kepler or TESS, as the transit depth of a faraway moon would be quite shallow. Also, the orbital period distribution implies that only several transits would have occurred during the Kepler mission, and barely 1 during a sector search with TESS. The best scenario for detecting ploonets would be through future microlensing missions such as Roman.}
    \label{fig:ploonetHist}
\end{figure}

The most likely outcome of our simulations is a ploonet (see Table \ref{tbl:results}). This is a multi-planet circumbinary system. There are two known multi-planet systems: Kepler-47 (3 transiting planets, \citealt{BHZ3}) and TOI-1338/BEBOP-1 (1 transiting and 1 non-transiting planet, \citealt{Kostov2020,Standing2023_bebop1}). These systems differ from our ploonets in three ways. First, the discovered planets are all much more massive than Earth, i.e., nothing is in the realm of a moon mass. Second, in both discovered multi-planet systems, the smallest planet is on the tightest orbit, which is opposite to our ploonets. Third, most of our ploonets have a much longer period than the discovered planets, owing to a very slow Type-I migration rate for moon-mass bodies. 


As mentioned in Sect.~\ref{sec:results_migtimescale}, over half of this population exists with orbital periods above roughly 1400 days (see Fig.~\ref{fig:ploonetHist}), which is much larger than all of the confirmed CBPs at the time of writing. This indicates that the current CBP demographics might be missing a large population of low-mass, long-period planets, such as the systems with ploonets in our simulation\footnote{This work only investigates single planet systems. When multiple host planets are present, the likelihood of a ploonet becoming unstable may be higher (see Sect.~\ref{sec:multipleplanets})}. If such systems exist in nature, they could potentially be probed by the upcoming Roman telescope. \citet{Penny_2019} shows that the design limit of the upcoming Roman microlensing survey is set to uncover bodies with masses as low as $0.02 M_{\oplus}$. This will allow the detection of Ganymede or Titan-sized objects to be within the realm of possibility, with the caveat that there is not a readily available method to discern between ploonets and planetesimals. 


\subsection{Habitable Zone Planets \& Moons}\label{sec:HZ}

Of the 14 currently known transiting CBPs, 5 orbit within the Habitable Zone (HZ) of their host binary \citep[see the introduction in][]{Kong_2022}. While the confirmed CBPs are gas giants and are unable to host life, their rocky satellites may be able to. Whilst it is thought that planets below 0.3 $M_{\oplus}$ cannot support atmospheric growth or plate tectonics on timescales required to harbor life \citep{Raymond_2007}, it may be possible that moons of gas giants can reach earth size or larger \citep{Teachey_2018}. Therefore, the occurrence rate of smoons has implications for the discussion of habitability in our simulated systems.

To model the HZ for a binary system, we follow the scheme laid out in \citet{Haghighipour_2013} with slight modifications. In that work, the authors derived the instantaneous habitable zone around binary stars by equating the total flux received by the planet to the flux that the Earth would receive from the Sun at the edge of our Sun's habitable zone. The total flux received by a planet can be written as
\begin{equation}
    F = W_\mathrm A \frac{L_\mathrm A}{r^2_{\mathrm p-\mathrm A}}+W_\mathrm B \frac{L_\mathrm B}{r^2_{\mathrm p-\mathrm B}} \ ,
\end{equation}
with $L_i$ representing the luminosity of each star and $r_{\mathrm p-i}$ representing the distance between the planet to each star at a given moment. $W_i$ represents the spectral weight factor, which encodes the interactions between the planetary atmosphere and the spectral energy distribution of stars with different spectral types. In contrast to \citet{Haghighipour_2013}, in which the primary star is assumed to be fixed, we calculated the habitable zone around the center of mass of the binary stars. 

Furthermore, instead of determining the instantaneous habitable zone by equating the flux at any given moment, we take the time average of the flux from both stars that the planet would receive over the binary stars' orbital period, and then compare it to the habitable zone around the Sun. In our work, we assume the narrow habitable zone boundaries, corresponding with the Runaway Greenhouse (inner) and Maximum Greenhouse (outer) conditions \citep{Kopparapu_2013}. An object is considered to be within the habitable zone if its final orbit receives a time-averaged flux lying within the inner and outer extremes of the narrow zone.







\begin{table*}
    \centering
    \caption{Percentages of smoons and ploonets that orbit within the habitable zone for each population of host planets. The migration statistics in rows 2 and 3 do not include collisions, as the orbital elements were not tracked beyond the collision. In all cases, having a fully migrated host planet increases the likelihood of the moon residing in the HZ. Moon-sized objects migrate at timescales much larger than 1 Myr, resulting in the habitable zone ploonet percentages being much smaller than smoons. Of the 578 smoons, 43 are hosted by small planets ($R_p \leq 3R_{\oplus}$) that lie within the HZ.}
    \begin{tabular}{cccc}
         \hline
         \textbf{Population}\quad & \textbf{Ploonets (\%)} \quad& \textbf{Smoons (\%)} \quad& \textbf{All Outcomes (\%)}\\
         \hline
         All Systems & 8/758 (1) & 102/578 (18) & 110/2000 (6) \\
         Fully Migrated & 7/466 (2) & 78/186 (42) & 85/664 (13) \\
         Partially Migrated & 1/292 ($<$\ 1) & 24/392 (6) & 25/686 (4) \\
    \label{tbl:HZperc}
    \end{tabular}
\end{table*}


Table \ref{tbl:HZperc} shows the relative percentages of ploonets, smoons, and planets that reside in the habitable zone when the disc dissipates. It is roughly $4\times$ more likely for a smoon to exist in the HZ than a ploonet ($7\times$ more likely for fully migrated systems), owing again to the slower migration rate of moon-mass objects once they have escaped their host planet. Across the whole population, $8\%$ of smoons and ploonets orbit in the HZ after 1 Myr of system evolution. Of the 578 smoons in our population, 410 are hosted by small planets ($R_p \leq 3R_{\oplus}$), 43 of which reside in the HZ. Although this is a small fraction (roughly $3\%$ of our total sample), this demonstrates the possibility of a small, rocky CBP in HZ that could host a moon, analogous to our own Earth-Moon system.

It is important to note that the percentage of smoons that exist within the Habitable Zone could be significantly altered on Gyr timescales due to secular effects such as tidal interactions with the host planet. This calculated relative abundance of smoons in the HZ should be taken as an upper bound, and is meant to reflect habitability scenarios in young planetary systems. The ploonet population would not be affected by the same secular effects as smoons, however, there are a variety of assumptions that would imply these values should be taken as an upper bound as well (i.e., single planet+moon systems, co-planar disc evolution, uncertain disc lifetimes, etc.)


\subsection{Free Floating ``Planets''}\label{sec:freefloating}

The Galaxy is believed to be populated by free-floating or ``rogue'' bodies that are not bound to a star. These could be planetary mass objects, of which some have been discovered by microlensing \citep{Sumi2011,ClantonGaudi}, or smaller asteroid-like objects like those seen recently passing through our Solar System (e.g. `Oumuamua, \citealt{Meech2017}).

Binary stars have been touted as a potential birth environment for free-floating planets. \citet{Smullen2016,Sutherland2016} demonstrated that for unstable circumbinary planets, an ejection was much more likely than a collision with either star. \citet{fitzmaurice2022} showed that migrating multi-planet circumbinary systems frequently eject the lower mass planet. \citet{Coleman2024} proposed using the velocity distribution of free-floating bodies to test if they were ejected from binaries (higher velocity) or single stars (lower velocity). 

In this paper, we have demonstrated another pathway that produces free-floating bodies. In our simulations, moons are naturally ejected if they escape the planet when the planet is reasonably close to the disc edge. However, this outcome was the rarest, occurring in only $1\%$ of our systems. Another mechanism that supports the idea of binaries producing FFPs may lie in the large number of ploonets within our population. Whilst these ploonets by definition survive our simulations, if there were another planet(s) in the system, then a low-mass ploonet would be susceptible to ejection \citep{fitzmaurice2022}. 

\subsection{Assumptions \& Caveats}\label{sec:caveats}

\subsubsection{Additional Planets}\label{sec:multipleplanets}

We assume that a stellar binary only hosts one fully formed planet-moon system at the beginning of the planetary migration. The planet's inward migration is unimpeded by any other planets. The addition of other planets, in particular more massive ones, would likely lead to increased instability \citep{martin2022runninggauntletsurvival}. Such destabilizing effects would be more pronounced on ploonets, most of which remain in a wide orbit exterior to their former host planet.

\subsubsection{Migration Timescales}

It is assumed that the planet and moon have already formed before the simulations have started, with no mass accretion occurring along the way. This assumption is quite common amongst simulations of planetary migration, both hydrodynamical \citep{PierensNelson2013} and $N$-body \citep{Rein2012,martin2022runninggauntletsurvival}. 

Our $N$-body simulation of migration also has many assumptions and simplifications. We use a fixed migration rate, calculated based on the \citet{lubow2010planetmigration} Lindblad torque at a somewhat arbitrary distance of 1 AU. Due to computational resource constraints, we do not model any effects of turbulence in the disc, e.g., with stochastic forcing, although this was not seen to significantly impact the results of circumbinary planet migration in \citet{martin2022runninggauntletsurvival}.

Finally, we assume that only the oblateness of a host planet and subsequent evection resonance impact the moon's orbit as a moon. We assume that tidal and spin interactions with the planet would occur over a longer timespan than the migration, and hence we neglect them. We leave it to other studies (similar to \citealt{KisareFabrycky2024}) to test if our smoons would remain long-term stable as the smoon evolves slowly post-migration.

We, like previous studies \citep{Namouni2010,Spalding_2016,Pu2025}, ignore disc-driven migration of the moon relative to the planet under the assumption that the gas left around the moon wouldn't be dense enough to affect the Moon (we provide an order of magnitude calculation in the Appendix for justification). However, if the moon's orbit is affected by this gas in any way, it could lead to significant instabilities resulting in its destruction/ejection. We encourage future studies that explicitly consider these effects, but further investigations are beyond the scope of this paper.

\subsubsection{Coplanarity}

All systems are presumed to be coplanar. This may be reasonable, given that the known CBPs all have orbits that are coplanar with the binary orbit to within $\sim 4^{\circ}$, although this might be an observational bias and not representative of the entire CBP population \citep{MartinTriaud2014,Li2016,Martin2017}. Misalignment of the planet and/or moon orbit could be destabilizing for the moon \citep{Hamers2018}, although we expect that during migration the protoplanetary disc would dampen inclinations. For reference, the orbital plane of the Moon is misaligned by $\sim 5^{\circ}$ relative to the Earth's orbit around the Sun. Misaligned moons could be discovered in transit, but their observational signature may pose additional complexities \citep{Kipping2009a,Kipping2009b,Martin2019Kepler1625}.

\subsubsection{Population Synthesis}

There is significant uncertainty in the simulated population. The binary stars are relatively well known, though this population likely oversamples secondaries with a mass of 0.19 $M_{\odot}$, as we set this if the selected secondary mass is below the stellar mass threshold.. For the planets, we have barely more than a dozen discoveries of CBPs, and with significant biases (e.g., it's harder to find longer period and smaller planets). In addition, we are simulating the initial location of the planets, which is even more uncertain. Finally, the simulated exomoon population is likely the least certain, as all relevant parameters are based on our own Solar System's moons and theoretical orbital distances based on Hill stability considerations. Ultimately, exomoon observations are required to make definitive statements about this population.

\subsubsection{Disc Lifetime}

Whilst there is an ongoing effort to rigorously constrain the lifetime of circumbinary protoplanetary discs \citep{binarydisklifetime, diskLifetime2, constrain_disc_lifetime}, we assume the lifetime of the protoplanetary disc is 1 Myr. For this reason, it is possible that some migrating bodies would not migrate to where they would if the disc had an infinite lifetime. This is especially true for moons, where the typical migration rate is on the order of 10-100 Myr. Thus, a fraction of our population will never reach a point where resonance overlap is a capable ejection mechanism.


\section{OUTLOOK \& CONCLUSION}\label{sec:conclusion}

Whilst there are currently no exomoon candidates around CBPs, our work shows that CBPs are capable of retaining moons through the migration process. Since CBPs are migrating within a truncated protoplanetary disc, they do not migrate as far as planets around single stars. This means that their Hill sphere does not shrink as much, which aids their ability to retain moons. This work shows that wide binaries are more likely to host CBPs with exomoons than tight binaries, as binaries separated by 0.3 AU host moons initially residing at roughly 15\% of their host planet's Hill radius, compared to 5\% for binaries separated by 0.1 AU. 

Our work also has implications for habitable worlds. For the known circumbinary planets, whilst they might be gas giants, $\sim 30\%$ of them reside within the habitable zone. Across all of our systems, 17\% of smoons and 4\% of ploonets exist within the habitable zone of their star systems after 1 Myr of disc evolution. Future population synthesis studies, along with successful CBP observations, will allow for better constraints on the occurrence rates of habitable zone objects across all star systems.

Finally, our work has implications for present and upcoming observing missions. The plausibility of habitable exomoons around CBPs may motivate JWST, CHEOPS, or similar searches for exomoons using transit and/or transit timing methods. We also show that Ganymede-mass objects can be ejected from CBP systems, but this situation is rare. Upcoming microlensing surveys, such as Roman, will be important in confirming the occurrence rates of these objects.

\section*{Acknowledgements}
We would like to thank various members of the astrophysics groups at Tufts University and The Ohio State University for valuable discussions that helped grow ideas within this paper. We thank the anonymous referee for thoroughly reviewing our manuscript, which greatly improved the quality of our work. We would also like to thank our families for their unwavering support throughout our work. This work commenced in Spring 2024 as a 3-credit research course taught at Tufts University: AST-192 ``Special Studies in Astrophysics''. All enrolled students contributed and are featured as co-authors.


\bibliography{sample631}{}
\bibliographystyle{aasjournal}


\appendix
\section*{Estimate of Residual Gas Within the Migrating Protoplanetary Disc}
\addcontentsline{toc}{section}{appendix}

In our simulations, the planet migrates according to our simple prescription of Type-I or Type-II migration. The moon, however, does not undergo any disc-driven migration relative to the planet. That is to say, it is effectively tied to the planet as the planet itself migrates. This approximation matches that of earlier studies by \citet{Namouni2010,Spalding_2016,Pu2025,Bolmont2025}. In addition, we consider that the moon does not experience drag from the surrounding gas. Here, we perform a rough calculation to test the validity of the assumption that the gas disc does not affect the moon.

In the context of the moon having disc-driven migration, there must be structured gas of sufficient density within the planet's Hill sphere. Consider a planet undergoing Type-I migration, i.e. the surrounding gas has not been substantially cleared out. 

We estimate the density of material within a planet's Radius based on the flux of disc material flowing onto the star. Consider a star to be accreting at $\dot{M}_\star = 10^{-7} M_{\odot} \ yr^{-1}$, a relatively high rate. As a rough approximation, take the star to be accreting off the disc within a radius $a_p$. From this, the fraction $f$ of the accreted flux that is within the planet's Hill radius comes from a ratio of areas:
\begin{equation}
    f = \biggl(\frac{R_{\rm H}}{a_{\rm p}}\biggr)^2 \ . 
\end{equation} 

\citet{Moldenhauer2022} find that the amount of time that material spends within the planetary Hill radius, referred to as the ``recycling time" $t_{\rm r}$, is on the same order as a dynamical timescale (e.g. $t_{ \rm r} \propto \Omega^{-1} = P/(2\pi))$. Therefore, the instantaneous amount of mass within the planetary Hill radius is

\begin{equation}
    M_{\mathrm{gas}} \sim f\dot{M}_\star t_{\rm r} \ .
    \label{eqn:gasmass}
\end{equation}

Using reasonable quantities from our simulated population, ($a_{\rm p} = 0.5 \ \mathrm{AU} $, $M_{\rm p} = 1 M_{\oplus}$, $M_{\rm \star, total} = 1 M_{\odot}$), and assuming all the gas lies in a sphere of radius $r_{\mathrm{gas}} = \mathrm{0.48}R_{\rm H}$, Eq.~\ref{eqn:gasmass} yields a gas density of $\rho_{\mathrm{gas}} = 1.3\times10^{-12} \  \mathrm{g} \ \mathrm{cm}^{-3}$. This value is 4-6 orders of magnitude less than gas densities found in typical circumplanetary disc midplanes \citep{Ayliffe2009} and therefore not enough to cause the moon to undergo any disc-driven migration. 

Moreover, even if this gas density was enough to induce migration, since $\tau_a \propto \frac{1}{\sum_d}$, then the moon would migrate so slowly that the effects would be negligible within the disc lifetime. To reiterate, this estimate is done under the assumption of Type-I migration by the host planet. If it were instead Type-II migration (Sec.~\ref{sec:t2migration}), our calculated densities would be reduced by a further factor of $\sim 100$, since the planet largely clears out the surrounding disc material. 


In the context of the planet going through Type-II migration, hydrodynamic simulations have shown that there is a possibility for gas to stream into the gap created in the Type-II migration regime, which would have a steep effect on the dynamics of small objects \citep[cited as $\sim 10$ km in][]{2023ASPC}. The moons in our study, modeled after Ganymede, are well above this value with radii of $\sim 2600$ km. 

In some instances, gas drag may become relevant for more massive bodies if the disc gas is slowed down to sub-Keplerian speeds \citep{Adams_2009}. Nonetheless, in these instances, the density of the gas surrounding the body is still comparable to the disc's density. In our scenario, we showed that any residual gas/dust that is present in a satellite's path exists in densities well below any threshold that will cause a meaningful dynamical effect. We nevertheless encourage future hydrodynamical simulations that incorporate disc-driven migration at the scale of both a planet and a moon.




\end{document}